\documentclass[journal=jacsat,manuscript=article]{achemso}
\usepackage{amsmath} 
\usepackage[version=3]{mhchem} 
\usepackage[utf8]{inputenc}
\usepackage[T1]{fontenc}
\usepackage{soul}
\usepackage{bm}
\usepackage{threeparttable}
\usepackage{color}
\usepackage{booktabs}
\usepackage{graphicx}
\usepackage{lscape}
\usepackage{multirow}
\usepackage{textcomp}
\usepackage[symbol]{footmisc}

\newcommand{\cm}{cm$^{-1}$}

\author{Chen Qu}
\email{szquchen@gmail.com}
\affiliation{Independent Researcher, Toronto, Ontario M9B0E3, Canada}
\author{Thomas C. Allison}
\affiliation{National Institute of Standards and Technology, 100 Bureau Drive, Gaithersburg, MD 20899, USA} 
\author{Apurba Nandi}
\affiliation{Department of Physics and Materials Science, University of Luxembourg, L-1511, Luxembourg City, Luxembourg.}
\email{apurba.nandi@uni.lu}
\affiliation{Dipartimento di Chimica, Universit\`{a} degli Studi di Milano, via Golgi 19, 20133 Milano, Italy}
\author{Paul L. Houston}
\affiliation{Department of Chemistry and Chemical Biology, Cornell University, Ithaca, New York
14853, USA and Department of Chemistry and Biochemistry, Georgia Institute of
Technology, Atlanta, Georgia 30332, USA}
\author{Qi Yu}
\affiliation{Department of Chemistry, Fudan University, Shanghai, 200438, P. R. China }
\author{Riccardo Conte}
\affiliation{Dipartimento di Chimica, Universit\`{a} Degli Studi di Milano, via Golgi 19, 20133 Milano, Italy}
\author{Joel M. Bowman}
\email{jmbowma@emory.edu}
\affiliation{Department of Chemistry and Cherry L. Emerson Center for Scientific Computation, Emory University, Atlanta, Georgia 30322, USA.}

\title[]{``Gold-Standard'' $\Delta$-Machine Learned and Transferable Potential for Linear Alkanes }

\abbreviations{PES,PIP}

\begin{document}


\newpage
\begin{abstract}
The conformational properties of linear alkanes, C$_n$H$_{2n+2}$, have been of intense interest for many years.  Experiments and corresponding electronic structure calculations were first reported in the mid-2000s and continue to the present time. These focus on the minimum chain length where the transition from the linear minimum to the hairpin minimum occurs.  We recently reported a transferable many-body permutationally invariant polynomial (MB-PIP) for linear alkanes using B3LYP electronic energies, which do not account for dispersion. Here we report a $\Delta$-ML approach to elevate this B3LYP-based and new PBE0+MBD MB-PIP potentials using PNO-LCCSD(T)-F12 energies.  The new $\Delta$-corrected potentials predict the difference in these minima accurately, compared to benchmark CCSD(T) results, over the range \ce{C12H28} to \ce{C28H58}. Vibrational power spectra are also reported for \ce{C14H30} and \ce{C30H62} using the uncorrected and $\Delta$-ML B3LYP. These new PIP-MB potentials for linear alkanes are the most accurate ones currently available and can be used in studies of properties of linear alkanes.
\end{abstract}

\newpage

Hydrocarbons are important in fuels, plastics, and other industrial products. 
Consequently, there have been many studies, not only of their chemistry, but also of their physical and molecular properties.  The conformational properties of linear alkanes, C$_n$H$_{2n+2}$, have been of particular interest for many years. Early studies of hydrocarbons have been reviewed by Patterson.\cite{Patterson2020Modern} Molecular mechanics models have been developed by Allinger,\cite{Allinger2010Molecular} and by Jorgenson,\cite{Thomas2006Conformation, Kaminski1994Free,Jorgensen2024OPLS} among others.  

\begin{figure}
    \centering
    \includegraphics[width=0.75\linewidth]{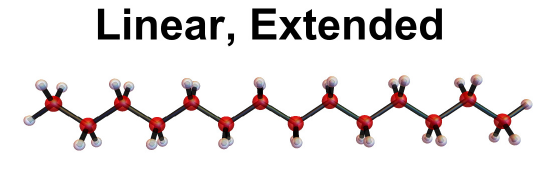}
    \includegraphics[width=0.75\linewidth]{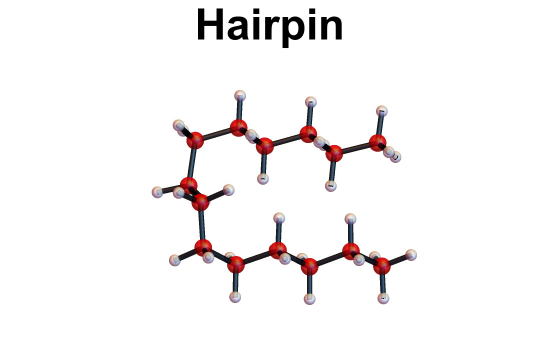}
    \caption{Linear extended and hairpin structures for the alkane \ce{C14H30}}
    \label{fig:exthairpin}
\end{figure}

Experiments and corresponding electronic structure calculations were reported in the mid-2000s, with a focus on the minimum chain length where the transition from the linear minimum to the hairpin minimum occurs.\cite{LuttschwagerSuhm2013,Byrd2014AtWhat,neese2015}. Theoretical electronic structure work continues to the present.  The paper by Liakos and Neese\cite{neese2015} is a recent and comprehensive study of the linear and hairpin minima using high-level \textit{ab initio} theory, namely  Domain Based Pair Natural Orbital Coupled Cluster theory (DLPNO-CCSD(T)). Rather than repeat the excellent background given there, we simply refer the interested reader to that paper. A major point of this careful study was the quantitative examination of the difference in geometry of these two minima and the resultant sensitivity of the energy difference.  This was examined for alkanes ranging from 
\ce{C6H14} to \ce{C18H38}, with the conclusion that the hairpin minimum drops below the ``extended'' linear one at \ce{C16H34} or \ce{C17H36}. This conclusion is in accord with earlier ones using lower-levels of electronic structure theory.\cite{LuttschwagerSuhm2013,Byrd2014AtWhat}  A final point about these ``single-point'' studies is that the importance of dispersion was noted by Bartlett and co-workers who wrote ``As the length of unbranched alkane chains reaches
some critical length, intramolecular dispersion forces
cause a self-solvation effect in which the chains assume a
folded conformation.''\cite{Byrd2014AtWhat} More recent, extended computational studies of linear alkane conformational energies by Ehlert et al.\cite{grimme2022} created the ACONFL dataset of energies.  Subsequent work using this dataset was reported by Santra and Martin\cite{martin2022} and most recently in 2023 by Werner and Hansen.\cite{PNO-LCCSD(T)-F12} The carefully benchmarked  PNO-LCCSD(T)-F12 approach used by Werner and Hansen is used here (see below for details). Figure \ref{fig:exthairpin} depicts the extended and hairpin conformers of alkanes, exemplified for \ce{C14H30}. 

Since 2023, there has been substantial progress in methodology in developing machine-learned potentials; for a recent review see ref. \citenum{jiangrev2025}.  With respect to linear alkanes, we recently reported permutationally invariant polynomial (PIP) machine-learned potentials for \ce{C14H30} from precise training of roughly 250 000 B3LYP electronic energies. One of those potentials was based on a novel many-body ``targeted" approach, denoted MB-PIP, that we showed is transferable to general linear alkanes.\cite{Chen2025Targeted} It was recently used in a study of the folding landscape.\cite{jcpc14h30} 

This MB-PIP transferable potential is the focus of work here. Specifically, we report a $\Delta$-ML approach to elevate the previous B3LYP-based and new PBE0+MBD MB-PIP potentials using 4515 PNO-LCCSD(T)-F12 energies.\cite{molpro2} The new $\Delta$-ML potentials provide energy differences between the hairpin and linear minima that are compared to directly calculated PNO-LCCSD(T)-F12 results for alkanes \ce{C12H26} to \ce{C28H58}.

The paper is organized as follows.  First, we give a brief review of many-body approaches in machine-learned potentials, including those based on PIP regression. Then, details of the approach taken for the linear alkanes are reviewed followed by details of the $\Delta$-ML approach. The results of the approach are given and assessed by comparing the energy difference between the hairpin and linear minima against the benchmark numbers. Power spectra of \ce{C14H30} are presented using B3LYP and $\Delta$-B3LYP Potentials at two total energies corresponding to temperatures of 10 and 300 K.

Many-body approaches to machine learned potentials of large molecules, clusters, and condensed phases are becoming the standard approach; however, the details vary greatly depending on the method. For a recent review including a list of more than 90 open-source software programs, see reference \citenum{jiangrev2025}. Among this large group of approaches, permutationally invariant polynomial regression has played an important role, both historically and also very recently.\cite{pippersp2025} This is reviewed briefly below which concludes with the details of a new $\Delta$-machine learning approach for our MB-PIP approach.  That approach is applied to two DFT-based MB-PIP potentials for linear alkanes. \cite{jctctwists,jpcac14h30,Chen2025Targeted}

Many-body representations of potentials are generically given by
\begin{equation}
  V(1,\cdots,N)=\sum_{i=1}^NV_{1-b}(i)+\sum_{i>j}^NV_{2-b}(i,j)+\sum_{i>j>k}^NV_{3-b}(i,j,k)+\sum_{i>j>k>l}^NV_{4-b}(i,j,k,l) + \cdots,
  \label{eq:mb1}
\end{equation}

\noindent This representation is well-known for non-covalent interactions, e.g., for water, where each ``body'' is a water monomer.\cite{WHBB1,mbpol3b,q_AQUA,qAQUApol,Mbpol23}.  In these widely-known potentials each $n$-body interaction is fit to gold-standard CCSD(T) energies using a basis of PIPs.  The most recent of these, q-AQUA,\cite{q_AQUA} q-AQUA-pol,\cite{qAQUApol} and MB-pol(2023)\cite{Mbpol23} fit up to 4-body interactions.

For covalent interactions, i.e., for a single (large) molecule, MB-PIP approaches are just appearing in the literature.  In this case, there isn't an obvious partitioning in terms of ``bodies''.  Paesani and co-workers recently suggested a physically based partitioning of linear hydrocarbons and developed a transferable PIP-based model.\cite{BullVulpe2023} Shortly after, we proposed and tested an atom-based PIP-approach that is different from -- albeit inspired by -- a PIP representation of an atom-centered expansion approach exemplified by work of Csanyi and co-workers\cite{aPIP2021}.  In our approach\cite{jctctwists,jpcac14h30} the monomers are atoms, and these interactions are enumerated for all atoms in the molecule.  For an $n$-body interaction among $N$ atoms, there are a total of $N!/[n!(N-n)!]$  interactions. To limit the number of interactions, physical range cutoffs are used for each $n$-body interaction, and details can be found in references \citenum{jctctwists,Chen2025Targeted} which report our first application to the 44-atom alkane \ce{C14H30} and transferability to alkanes as large as \ce{C30H62}, respectively. In this example (and also for other two-element molecules, clusters, materials, etc.,) there are three types of 2-b interactions, i.e.,  CC, HH, and CH interactions, four types of 3-b, and five types of 4-b interactions. (We do not consider 5-b interactions.) For each interaction beyond 2-b, an appropriate PIP basis is used and there are fewer of these than the types of interactions.  To see this, consider the 3-b CCC and HHH interactions.  While these are different, the PIP basis is the same, namely the basis for $A_{3}$ but the corresponding sets of linear coefficients are different. The full expression, up to $V_{4-b}$, for this MB-PIP representation has been given previously;\cite{Chen2025Targeted} here we just illustrate this for $V_{3-b}$.  
\begin{equation}
\label{eq:mb3b}
\begin{aligned}
V_{3-b}=
&\sum_{\text{C}_i,\text{C}_j,\text{C}_k}V^{(\text{CCC})}_{3-b}(y_{ij},y_{ik},y_{jk})
+\sum_{\text{C}_i,\text{C}_j,\text{H}_k}V^{(\text{CCH})}_{3-b}(y_{ij},y_{ik},y_{jk})+\\
&\sum_{\text{C}_i,\text{H}_j,\text{H}_k}V^{(\text{CHH})}_{3-b}(y_{ij},y_{ik},y_{jk})
+\sum_{\text{H}_i,\text{H}_j,\text{H}_k}V^{(\text{HHH})}_{3-b}(y_{ij},y_{ik},y_{jk})\\
=&\sum_{m}\sum_{\text{C}_i,\text{C}_j,\text{C}_k} c_m^{(\text{CCC})}p_m^{(A_3)}(y_{ij}, y_{ik}, y_{jk})
 +\sum_{m}\sum_{\text{C}_i,\text{C}_j,\text{H}_k} c_m^{(\text{CCH})}p_m^{(A_2B)}(y_{ij}, y_{ik}, y_{jk})+\\
 &\sum_{m}\sum_{\text{C}_i,\text{H}_j,\text{H}_k} c_m^{(\text{CHH})}p_m^{(A_2B)}(y_{jk}, y_{ij}, y_{ik})
 +\sum_{m}\sum_{\text{H}_i,\text{H}_j,\text{H}_k} c_m^{(\text{HHH})}p_m^{(A_3)}(y_{ij}, y_{ik}, y_{jk}),
\end{aligned}
\end{equation}
where now the permutational symmetry of the two $3$-b PIPs are $A_{3}$ and $A_2B$.  

For the expansion up to 4-b terms, the general permutational symmetry of the PIPs bases are $A_4$, $A_3B$, and $A_2B_2$. Also, we note that the distinct 2-b, 3-b, and 4-b PIP bases can have different range parameters for the corresponding Morse variables, and different maximum polynomial orders. Once these bases are set up, the linear coefficients are determined using a standard over-determined linear least-squares fit to the total energies of this molecule. 

As shown previously, this MB-PIP is transferable to linear alkanes and this was demonstrated for alkanes with as many as 30 carbon atoms.\cite{Chen_Houston_Bowman_vinylidene_2010} We also note that the MB-PIP approach is to be distinguished from fragmentation representations\cite{collins_rev} and the use of a fragmented (PIP) basis.\cite{NandiBowman2019}  This approach was used successfully to represent the PES for \ce{C14H30}.\cite{jctctwists}

This MB-PIP potential was fit to a dataset of roughly 250 000 B3LYP energies for \ce{C14H30}; details of the dataset and fitting are given elsewhere\cite {jctctwists,Chen2025Targeted}. We note here and also in a very recent paper that dispersion is not accounted for in these straight B3LYP energies. So the central objective of the work here is to correct this in a single step by using $\Delta$-ML at the CCSD(T) level.  In addition, the above expression for the MB-PIP is employed here for a new set of PBE0+MBD energies \cite{PBE0_1,PBE0_2,mbd2012}, computed with the ``intermediate'' basis settings of the FHI-aims electronic structure package\cite{FHI_aims,FHI_aims_2}, for the same dataset of configurations. Since this approach does contain dispersion, $\Delta$-ML applied to this PES is not as heavy a ``lift''. 


The $\Delta$-ML approach is given by the general equation\cite{NandiDeltaML2021,JCTCPerspDelta}
\begin{equation} 
\label{eq:1}
    V_{LL{\rightarrow}CC}=V_{LL}+\Delta{V_{CC-LL}},
\end{equation}
where $V_{LL{\rightarrow}CC}$ is the corrected PES,  $V_{LL}$ is a PES fit to low-level DFT electronic data, and $\Delta{V_{CC-LL}}$ is the correction PES based on high-level (usually coupled cluster) energies.  In the present context $V_{LL}$ is the above MB-PIP fit to B3LYP or PBE0+MBD energies and $\Delta{V_{CC-LL}}$ is the analogous fit to the difference between high-level and DFT energies.  Here we apply PNO-LCCSD(T)-F12 energies, implemented in Molpro,\cite{MOLPRO2023}\footnote[2]{Identification of certain commercial equipment, instruments, software, or materials does not imply recommendation or endorsement by the National Institute of Standards and Technology, nor does it imply that the products identified are necessarily the best available for the purpose.} to obtain the correction term $\Delta{V_{CC-LL}}$. 

In general, the expectation is that the 4-b interactions are shorter-ranged than 3-b than 2-b, so we applied smaller Morse range parameter and cutoff distance for 4-b, and larger values for 2-b. The 1-b energy is simply the sum of the energies of isolated carbon and hydrogen atoms computed using B3LYP/cc-pVDZ theory and it is not a trainable parameter.  For the $2$-b,  a single Morse range parameter of 2.5 bohr was used, with a maximum power of 10, so the total number of $2$-b coefficients is 30. The cutoff distance for $2$-b is 18.0 bohr. For the 3-b, the Morse range parameter is 1.8 bohr, and the maximum polynomial order of the PIPs is 8.  The switching function is applied when the maximum distance in a trimer is between 12.3 and 14.2 bohr, and the energy contribution is 0 when the maximum internuclear distance in the trimer is beyond 14.2 bohr. For the $4$-b bases, the Morse range parameter is 1.2 bohr, and the maximum polynomial order is 6. The switching range for $4$-body is between 8.5 and 10.4 bohr. All the bases are purified, \cite{purified13, purified14} that is, any polynomial that does not go to zero when an atom is infinitely far away from the remaining atoms is removed from the bases. Using the polynomial orders mentioned above, there are 32, 78, 78, 32, 40, 115, 174, 115, and 40 PIPs for CCC, CCH, CHH, HHH, CCCC, CCCH, CCHH, CHHH, HHHH, respectively. These sum up to 734 undetermined linear coefficients, including the 30 from all the 2-b interactions. 

These coefficients were determined in a single least-squares fit to \textit{ab initio} energies of \ce{C14H30}. (See ref. \citenum{h3o22024} for details of the least-squares implementation.) The training (on \ce{C14H30}) gives a root-mean-square error on energies of 115 \cm~ (0.33 kcal/mol), and corresponds to an error of 2.6 \cm~ (0.007 kcal/mol) per atom. This is very small precision error compared to typical ML values for such a large molecule and dataset.  Even though it should be clear, \textit{prima facie}, that these many-body terms, trained on \ce{C14H30}, are transferable to all alkanes.  The precision of such transferability needs to be verified of course.  This was done by us where excellent precision was shown over the range \ce{C5H12} to \ce{C30H62}.\cite{Chen2025Targeted}

For $\Delta$$V_{CC-LL}$, a corresponding MB-PIP fit was done to 4514 energy differences between PNO-LCCSD(T)-F12b/AVTZ' energies (where AVTZ' means cc-pVTZ basis for H and aug-cc-pVTZ for C) and B3LYP or PBE0+MBD ones for \ce{C14H30}. The 4514 configurations were chosen as follows: A first set of 1359 configurations comes from molecular dynamics simulations. Specifically, along the minimum energy path between the linear and hairpin structure, 7 additional local minima were located on the low-level PES. Molecular dynamics calculations (NVE) starting from these 9 stationary points (7 plus the linear and hairpin) were carried out and 151 configurations were collected from each trajectory. The second set of 459 configurations was obtained by assigning random displacement of coordinates to the 9 stationary points, 51 configurations each. The last 2544 configurations were randomly picked from the data set used for the low-level PES. For all the 4514 configurations, PNO-LCCSD(T)-F12b energies were obtained. The fitting also used the MB-PIP approach, with the same hyperparameters as the low-level PES, except the polynomial order. The maximum polynomial orders of 2-b, 3-b, and 4-b of the $\Delta$$V_{CC-LL}$ are 6, 7, and 5, respectively, resulting in 383 unknown coefficients. This very small number results in a trivial linear least-squares problem. The fitting errors are 12 \cm~ for the $\Delta$-B3LYP data and 7 \cm~ for the $\Delta$-PBE0+MBD data; these indicate very precise fitting. These small fitting errors are expected because the range of the energy difference is much smaller, specifically, the energy differences between PNO-LCCSD(T)-F12 and B3LYP span a range of 12 966 \cm, and the differences between PNO-LCCSD(T)-F12 and PBE0+MBD span only 3904 \cm.  We note that the larger range for the former than the latter is not surprising, because the B3LYP energies are dispersionless whereas the  PBE0+MBD ones describe dispersion accurately.  That the precision of the two correction PESs, $\Delta$$V_{CC-LL}$, are about the same is certainly a gratifying outcome of our machine learning approach.

To sum up thus far, we have obtained transferable low-level MB-PIP PESs based on B3LYP (with no dispersion) and PBE0+MBD energies, $V_{LL}$,  and corresponding correction PESs $\Delta$$V_{CC-LL}$, all trained on \ce{C14H30}. We now examine the performance of the transferability of the corrected PES, $V_{LL{\rightarrow}CC}$ for the determination of the energy difference between the hairpin and extended conformations over the range \ce{C12H26} to \ce{C28H58}.  For consistency, we use stationary configurations for these alkanes, obtained from direct optimization of the relatively efficient and (as seen below) accurate PBE+MBD method.

\begin{figure}
    \centering
    \includegraphics[width=0.85\linewidth]{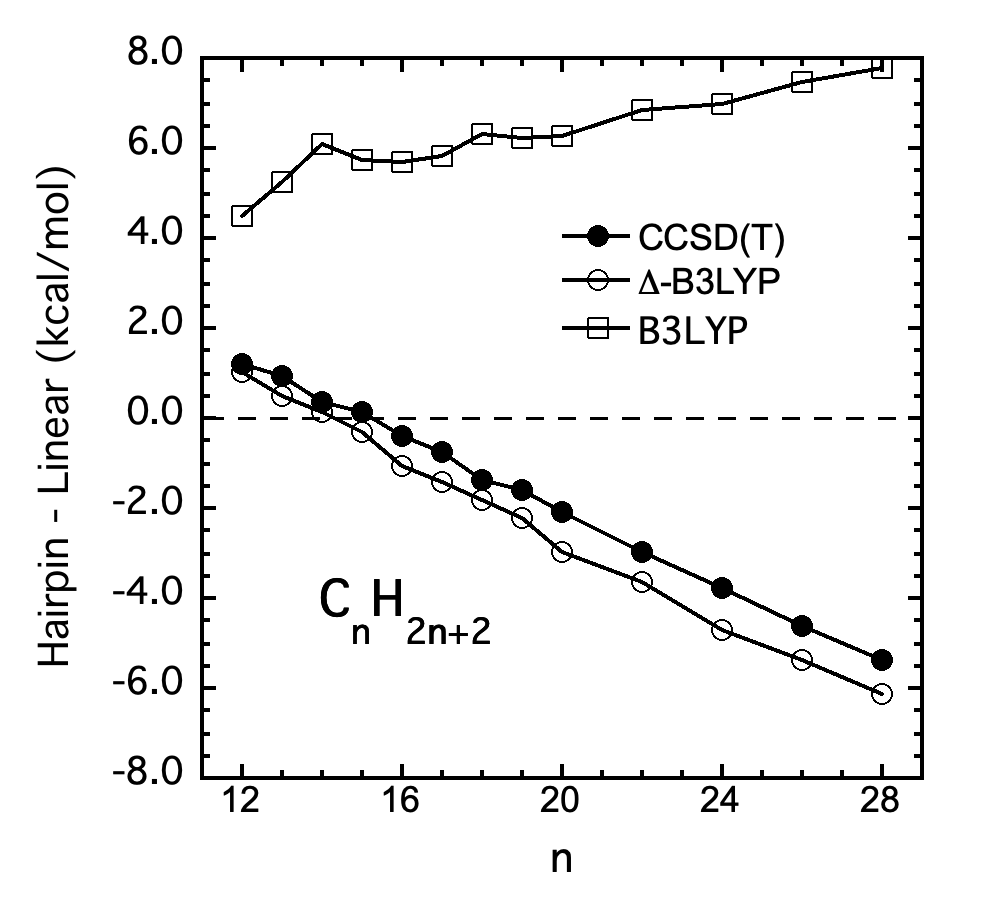}
    \caption{Energy difference between hairpin and linear minima from indicated sources and where ``CCSD(T)'' refers to PNO-LCCSD(T)-F12}
    \label{fig:B3LYPs}
\end{figure}

To begin we show results for the B3LYP and B3LYP- corrected PES shown graphically in Fig. \ref{fig:B3LYPs}. To be clear the ``B3LYP'' results are from the transferable MB-PIP B3LYP PES ($V_{LL}$) and those labeled  $\Delta$-B3LYP are from the transferable MB-PIP corrected B3LYP PES ($V_{LL{\rightarrow}CC}$). As seen, there is a major improvement of the B3LYP PES (with no dispersion) such that the corrected PES is very close to the benchmark CCSD(T) results.  The correction is not perfect, as expected, with a gap of roughly 0.5 kcal/mol.  This can be traced mainly to the fact that $\Delta$$V_{CC-LL}$ was trained on the difference between direct B3LYP and CCSD(T) energies and not the difference between the B3LYP PES and CCSD(T) energies.   Across the range of alkanes shown the mean absolute error of the PES is 0.5 kcal/mol

\begin{figure}
    \centering
    \includegraphics[width=0.85\linewidth]{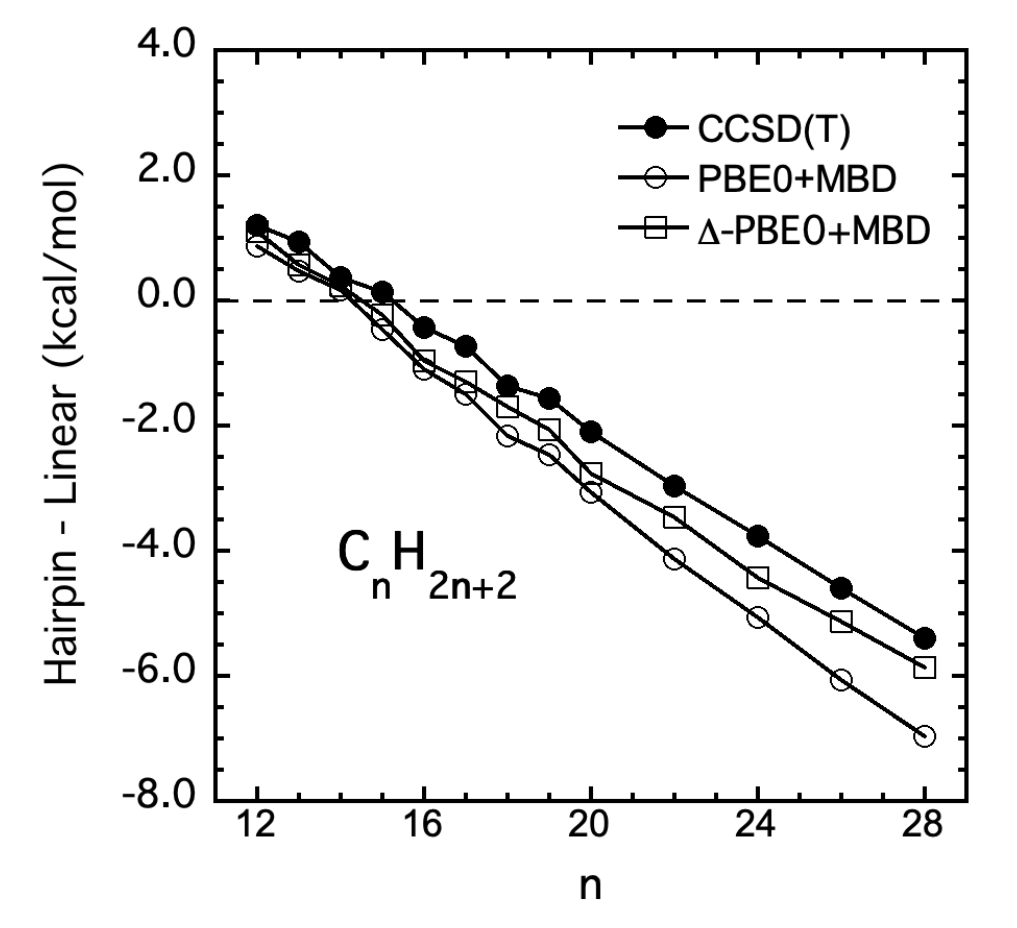}
    \caption{Energy difference between hairpin and linear minima from indicated sources and where ``CCSD(T)'' refers to PNO-LCCSD(T)-F12}
    \label{fig:PBE0+MBD}
\end{figure}

Next, consider the performance of the PBE0+MBD MB-PIP and corrected PBE0+MBD, MB-PIP PESs for these energy differences.  These are shown in Fig. \ref{fig:PBE0+MBD}. As seen the PBE0+MBD results are in good agreement with the CCSD(T), with differences of roughly 1-1.5 kcal/mol.  The corrected PBE0+MBD PES differences are reduced substantially to about 0.5 kcal/mol; essentially the same difference at seen for the corrected B3LYP MB-PIP PES.  The numerical values shown in Figures \ref{fig:B3LYPs} and \ref{fig:PBE0+MBD} are given in Table. \ref{tab:energy diffs}. These results indicate that the new transferable corrected MB-PESs based on B3LYP and PBE+MBD have successfully been elevated to the ``gold standard'' level, especially the previous B3LYP one.



    \label{fig:DeltaDFT}

\begin{table}[htbp!]
    \centering
    \begin{tabular}{cccccc}
        \hline
        $n$ & CCSD(T) & B3LYP & $\Delta$-B3LYP & PBE0+MBD & $\Delta$-PBE0+MBD \\
        \hline
        12 & 1.20 & 4.50 & 1.00 & 0.92 & 1.10 \\
        13 & 0.92 & 5.24 & 0.50 & 0.46 & 0.58 \\
        14 & 0.37 & 6.07 & 0.14 & -0.10 & 0.22 \\
        15 & 0.12 & 5.75 & -0.33 & -0.52 & -0.25 \\
        16 & -0.42 & 5.69 & -1.08 & -1.24 & -0.98 \\
        17 & -0.75 & 5.82 & -1.42 & -1.67 & -1.29 \\
        18 & -1.36 & 6.30 & -1.83 & -2.29 & -1.71 \\
        19 & -1.58 & 6.23 & -2.20 & -2.67 & -2.06 \\
        20 & -2.09 & 6.25 & -2.96 & -3.34 & -2.77 \\
        22 & -2.98 & 6.85 & -3.65 & -4.26 & -3.45 \\
        24 & -3.77 & 6.96 & -4.70 & -5.31 & -4.43 \\
        26 & -4.61 & 7.45 & -5.39 & -6.07 & -5.12 \\
        28 & -5.39 & 7.78 & -6.15 & -6.88 & -5.86 \\
        \hline
    \end{tabular}
    \caption{Energy difference (kcal/mol) between hairpin and linear minima from indicated calculations, and where ``CCSD(T)'' refers to direct PNO-LCCSD(T)-F12 calculations, and all other values were obtained using the corresponding MB-PIP PES. These are all evaluated at the PBE0+MBD optimized configurations.}
    \label{tab:energy diffs}
\end{table}


\newpage
There are numerous studies that can be done with these corrected MB PESs.  One that we present here are calculations of the vibrational power spectra of \ce{C14H30} from molecular dynamics simulations, using the B3LYP PES and $\Delta$-B3LYP PES. These were performed as we did recently,\cite{jpcac14h30} by running a classical trajectory initiated from a specific configuration; the final spectra are averages for five trajectories, run for approximately 1 ps, at each temperature and potential. Here we considered both the linear and hairpin minima.  The trajectories were run using the NVE protocol and for zero total angular momentum using our in-house software. The total energy is obtained from the simple correspondence between the average energy <E> and temperature $T$ for classical harmonic oscillators. (For \ce{C14H30} <E> = 126RT.) Temperatures of 10 and 300 K were considered with the goal of examining both the difference in the spectra at these two temperatures and also the differences between the B3LYP PES and $\Delta$-B3LYP PES.
The power spectra were obtained by using an open source web platform,\cite{web-platform, web-platform_b} recently developed at the University of Milan, which is based on the time-averaged Fourier transform of the Cartesian velocity autocorrelation function.\cite{web-platform_art2}
 
We examined such trajectories in detail previously,\cite{jpcac14h30} and noted that at 300 K many conformations are visited and so the final spectra are virtually identical whether starting from the hairpin or linear minimum.  However, at 10 K this is not the case, and the spectra obtained are more representative of the harmonic normal mode spectra at these minima.  With this in mind consider the 10 K results first.  There are clear differences between the hairpin and linear spectra in the range 0 to 1000 cm$^{-1}$ and this is seen for both the B3LYP and $\Delta$-B3LYP potentials. And, not surprisingly, we also see significant differences between these potentials in this spectral range.  These differences are largely quenched for the 300 K spectra for both potentials.  So, one conclusion from these results is that signatures of the effects of dispersion are present at 10 K in the spectral range 0 to 1000 cm$^{-1}$.  This conclusion is consistent with the double harmonic analysis done by Luttschwager and Suhm in 2013. \cite{LuttschwagerSuhm2013} 

A second, and again not surprising, conclusion from these spectra is that the high-frequency CH-stretch region is largely insensitive to temperature and potential.

\begin{figure}
    \centering
    \includegraphics[width=0.9\linewidth]{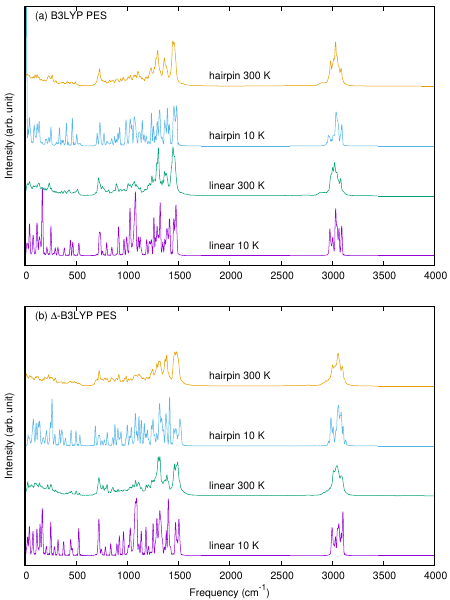}
    \caption{Power Spectra of \ce{C14H30} using B3LYP and $\Delta$-B3LYP PES at indicated temperatures.}
    \label{fig:power_spec}
\end{figure}


\newpage

The present results demonstrate that it is straightforward to elevate a low-level, e.g., DFT, MB-PIP potential for a large class of molecules, e.g., linear alkanes, to the CCSD(T) level.  This was done here for MB-PIP architecture, where all terms are fit ``at once''.  That the method is successful in elevating a low-level PES based on dispersion-less B3LYP calculations is very encouraging, albeit not necessarily surprising.  Note the correction to the low-level PES was done by fitting a difference potential using the ``at once'' approach.  The high-level energies were obtained for \ce{C14H30} using efficient and certified PNO-LCCSD(T)-F12 methods in Molpro 2023.  Also, we note that correction terms could be obtained using a smaller alkane.  This is something to be investigated in the future.   

The success was explicitly demonstrated for the energy difference between the hairpin and linear conformers of alkanes ranging from 12 to 28 carbon atoms against CCSD(T) energies.  As noted above the most recent ``single-point'' study of alkanes considered datasets  of 13, 18, and 22 configurations for alkanes with 12, 16, and 20 carbon atoms.\cite{grimme2022}  The present work considered 4514 configurations, spanning a large energy range,  at which PNO-LCCSD(T)-F12 energies were obtained. These were used to precisely fit a MB-PIP difference potential with precision of less than 0.03 kcal/mol.  

Finally, we note a different, ``building block'', approach was recently introduced by Paesani and co-workers for polyalanine chains.\cite{deltapaesani2025}  This work uses and builds on the many-body approach based on the selection of monomers, ``chemically intuitive building blocks.''\cite{BullVulpe2023}  In the more recent paper on polyalanine, each building block is separately elevated to the CCSD(T) level (using the DLPNO-CCSD(T) method). This is certainly a reasonable approach.  However, a statement made in that paper about our MB-PIP approach is contradicted by the present results.  Specifically, it was stated that  ``...the `many-body strategy' builds PIP terms for interactions involving sets of 2, 3, or 4 atoms, which are then summed to predict the total molecular energy.... because these models require whole-molecule reference data for training, they are typically limited to DFT-level calculations due to the prohibitive cost of higher-level quantum methods.'' The present work demonstrates that the MB-PIP approach can extend to higher-level methods using a standard $\Delta$-ML approach and without the need to select molecule-specific building blocks.

To summarize, the present paper reports two significant results. One is the successful elevation of DFT-based, transferable MB-PIP potentials for alkanes to the CCSD(T) level. Second, we demonstrated quantitatively the major and increasing importance of dispersion in determining the folding conformation of linear alkanes.

This new transferable potential can now be used in many applications and tests of other potentials for this important class of molecules. For example, our most work was a study of energies and configurations of tens of thousands of local minima and first order saddle-points of \ce{C14H30} using the B3LYP MB-PIP PES.\cite{jcpc14h30} It is now possible to conduct that study for a MB-PIP PES that includes dispersion and also for a range of alkanes. Also, the new MB-PIP PES can be used to test the new generation of universal machine-learned ones, such as MACE-OFF,\cite{mace-off} SO3LR,\cite{so3lr_2025} ANI,\cite{ANI19} Allegro\cite{allegro} and OMNI-P1.\cite{OMNI-P1}  






\begin{acknowledgement}
We thank Andreas Hansen for advice on the PNO-LCCSD(T)-F12 calculations using Molpro. J.M.B. thanks NASA, grant 80NSSC22K1167, for financial support. 
A.N. thanks Prof. Alexandre Tkatchenko for the financial support from PHANTASTIC grant INTER/MERA22/ 16521502/PHANTASTIC. R.C. thanks Universit\`a degli Studi di Milano for financial support through grant PSR2024 - linea 2.

\end{acknowledgement}

\section{Data Availability}
The PNO-LCCSD(T)-F12 electronic energies used in this paper are available at\\
https://github.com/jmbowma/QM-22






\bibliography{refs}

\end{document}